\documentclass[10pt,twocolumn,twoside]{IEEEtran}
\usepackage{amsmath}
\usepackage{amssymb}
\DeclareMathOperator*{\argmax}{arg\,max}
\DeclareMathOperator*{\argmin}{arg\,min}

\usepackage{graphicx}
\usepackage{epstopdf}
\usepackage{amsmath}
\usepackage{array}
\newcommand{\PreserveBackslash}[1]{\let\temp=\\#1\let\\=\temp}
\newcolumntype{C}[1]{>{\PreserveBackslash\centering}p{#1}}
\newcolumntype{R}[1]{>{\PreserveBackslash\raggedleft}p{#1}}
\newcolumntype{L}[1]{>{\PreserveBackslash\raggedright}p{#1}}
\usepackage[usenames]{color}
\usepackage{colortbl,booktabs}
\usepackage{tabularx}
\usepackage{multicol}
\usepackage{booktabs}
\usepackage[Symbol]{upgreek}
\usepackage{bm}
\usepackage{cite}
\usepackage{balance}
\usepackage[boxed]{algorithm2e}
\usepackage{mathrsfs}
\usepackage{url}

\usepackage{threeparttable}
\usepackage{amsmath}
\usepackage{dcolumn}
\usepackage{multirow}
\usepackage{booktabs}
\usepackage{amsfonts}

\newcolumntype{d}[1]{D{.}{.}{#1}}

\begin{document}
\title{Codebook Design for Channel Feedback in Lens-Based Millimeter-Wave Massive MIMO Systems}
%
\author{\IEEEauthorblockN{Author list}\vspace{-0mm}}
\author{Wenqian Shen,~\IEEEmembership{Student Member,~IEEE},
Linglong Dai,~\IEEEmembership{Senior Member,~IEEE}, Yang Yang,~\IEEEmembership{Member,~IEEE}, Yue Li,~\IEEEmembership{Member,~IEEE}, and Zhaocheng Wang,~\IEEEmembership{Senior Member,~IEEE}
\thanks{All authors are with the Department of Electronic Engineering, Tsinghua University, Beijing 100084, China (E-mails: swq13@mails.tsinghua.edu.cn;
 \{daill, yy0553, lyee, zcwang\}@tsinghua.edu.cn).}
\thanks{This work was supported by the National Key Basic Research Program of China (Grant No. 2013CB329203),  the National Natural Science Foundation of China (Grant Nos. 61571270 and 61571267),  the Beijing Natural Science Foundation (Grant No. 4142027), and the Foundation of Shenzhen government.}
}
\maketitle
\begin{abstract}
The number of radio frequency (RF) chains can be reduced through beam selection in lens-based millimeter-wave (mmWave) massive MIMO systems, where the equivalent channel between RF chains and multiple users is required at the BS to achieve the multi-user multiplexing gain.  
However, to the best of our knowledge, there is no dedicated codebook for the equivalent channel feedback in such systems.
In this paper,
we propose the dimension-reduced subspace codebook,
which achieves a significant reduction of the feedback overhead and codebook size.
Specifically, we firstly utilize the limited scattering property of mmWave channels to generate the high-dimensional vectors in the channel subspace.
Then, according to the function of lens and beam selector,
we propose the dimension-reduced subspace codebook to quantize the equivalent channel vector.
Moreover, the performance analysis of the proposed codebook is also provided.
Finally, simulation results show the superior performance of the proposed dimension-reduced subspace codebook compared with conventional codebooks.
\end{abstract}

\IEEEpeerreviewmaketitle

\section{Introduction}\label{S1}
\IEEEPARstart{M}{illimeter}-Wave (MmWave) massive MIMO has been regarded as a promising key technique for 5G due to the large available bandwidth and high spectrum efficiency \cite{JTSP_HRobert_OverviewMmwave}.
One key challenge for practical mmWave massive MIMO systems, where each antenna is usually equipped with a radio frequency (RF) chain, is the unaffordable hardware cost and energy consumption caused by a large number of RF chains.
To reduce the number of RF chains, the new concept of lens-based mmWave massive MIMO has been recently proposed \cite{TAP_JBrady_Beamspace,ICASSP_GSong_BeamspaceTransceivers,JSAC_YZeng_Electromagnetic,TCOM_YZeng_PDMA}.
By utilizing the energy focusing property of lens, mmWave signals from different directions are concentrated on different antennas, so that the traditional spatial channel is transformed into beamspace channel.
Due to the severe path loss of mmWave signals, the number of effective channel paths is limited \cite{Access_TSRappaport_mmwave}.
As a result, the beamspace channel is sparse. 
Then, using a beam selector, the number of RF chains can be significantly reduced, so the dimensions of equivalent channels can be also reduced.

The dimension-reduced channel state information (CSI) between RF chains and users, which is called as \textit{equivalent CSI}, is required at the BS to perform channel adaptive techniques such as precoding and power allocation.
However, to the best of our knowledge, there is no dedicated codebook to realize the equivalent CSI feedback in the emerging lens-based mmWave massive MIMO systems.
The classical Grassmannian codebook \cite{TIT_DJLove_Grassmannian} and random vector quantization (RVQ)-based codebook \cite{TIT_NJindal_MIMOBroadcast} are not directly applicable here, since the dimension-reduced equivalent CSI in lens-based mmWave massive MIMO systems is not Rayleigh distributed.

In this paper, to fill in this gap, we propose the dimension-reduced subspace codebook for equivalent CSI feedback in lens-based mmWave massive MIMO systems\footnote{Simulation codes are provided to reproduce the results presented in this paper: \url{http://oa.ee.tsinghua.edu.cn/dailinglong/publications/publications.html}.}.
Firstly, we utilize the limited scattering property of mmWave channels to generate the high-dimensional vectors in the channel subspace,
which depends on the angle-of-departure (AoD) of channel path.
Then, according to the function of lens and beam selector,
we propose the dimension-reduced subspace codebook to quantize the equivalent channel vector using less bits than traditional codebooks.
Moreover, we provide the performance analysis of the proposed codebook, which is verified by simulation results.

\emph{Notation}:
Lower-case and upper-case boldface letters denote vectors and matrices, respectively;
$(\cdot)^{\rm{T}}$, $(\cdot)^{\rm{H}}$ and $(\cdot)^{-1}$ denote the transpose, conjugate transpose, and inverse of a matrix, respectively;
$\mathbf{\Phi}^{\dagger}=\mathbf{\Phi}(\mathbf{\Phi}^{\rm{H}}\mathbf{\Phi})^{-1}$ is the Moore-Penrose pseudo-inverse; $\rm{E}[\cdot]$ denotes the expectation operator;
$\mathbf{I}_P$ denotes the identity matrix of size $P\times P$.

\section{System Model}\label{S2}
In this section, we briefly introduce the typical mmWave massive MIMO channel model at first.
Then, we present the equivalent channel in lens-based mmWave massive MIMO systems.
\subsection{MmWave Massive MIMO Channel Model}\label{S2.1}
 We consider a mmWave massive MIMO system with $M$ antennas at the base station (BS) and $K$ single-antenna users ($M\gg K$) in this paper.
 We adopt the widely used ray-based mmWave channel model \cite{JTSP_HRobert_OverviewMmwave},
 where the downlink channel vector $\mathbf{h}_k\in\mathbb{C}^{M\times 1}$ between the BS and the $k$-th user can be described as
 \begin{align}\label{eq_hk1} 
 \mathbf{h}_k=\sum_i^{P_k}g_{k,i}\mathbf{a}(\psi_{k,i}),
 \end{align}
 where $P_k$ is the number of resolvable paths from the BS to the $k$-th user,
 $g_{k,i}$ is the complex gain of the $i$-th path of the $k$-th user, which is identically and independently distributed (i.i.d.) with zero mean and unit variance,
 and $\mathbf{a}(\psi_{k,i})\in\mathbb{C}^{M\times 1}$ is the BS antenna response of the $i$-th path of the $k$-th user.
  We consider the widely used uniform linear array (ULA) at the BS \cite{JTSP_HRobert_OverviewMmwave}, so $\mathbf{a}(\psi_{k,i})$ can be expressed as
 \begin{align}\label{eq_a}
 \mathbf{a}(\psi_{k,i})=\frac{1}{\sqrt{M}}\left[ 1,e^{-j2\pi\psi_{k,i}},\cdots, e^{-j2\pi\psi_{k,i}(M-1)}\right] ^{\rm{T}},
 \end{align}
 where $\psi_{k,i}=\frac{d}{\lambda}\sin(\theta_{k,i})$ with $\theta_{k,i}$ denoting the AoD of the $i$-th path of the $k$-th user,
 $d$ denoting the distance between BS antennas,
 and $\lambda$ denoting the wavelength of the carrier frequency.
 Note that (\ref{eq_hk1}) can be also expressed in the matrix-vector form as
  \begin{align}\label{eq_hk2}
 \mathbf{h}_k=\mathbf{A}_k\mathbf{g}_k,
 \end{align}
 where $\mathbf{g}_k=[g_{k,1},g_{k,2},\cdots,g_{k,P_k}]^{\rm{T}}\in\mathbb{C}^{P_k\times 1}$, and $\mathbf{A}_k=[\mathbf{a}(\psi_{k,1}),\mathbf{a}(\psi_{k,2}),\cdots,\mathbf{a}(\psi_{k,P_k})] \in\mathbb{C}^{M\times P_k}$.

\subsection{Equivalent Channel in Lens-Based MmWave Massive MIMO}\label{S2.2}
\begin{figure}[t]
\vspace{-3mm}
\center{\includegraphics[width=0.5\textwidth]{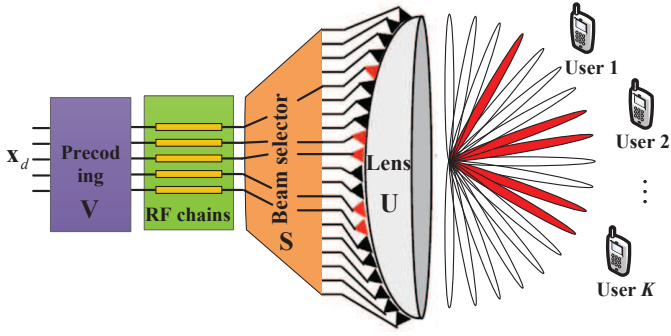}}
\vspace{-3mm}
\caption{Lens-based mmWave massive MIMO systems.}
\label{beamspaceMIMO}
\vspace{-1mm}
\end{figure}
The spatial channel $\mathbf{h}_k$ in (\ref{eq_hk2}) can be transformed into the beamspace channel through a carefully designed lens which performs as a phase shifter \cite{TAP_JBrady_Beamspace}. 
Specifically, as shown in Fig. \ref{beamspaceMIMO}, such lens can be characterized as a spatial discrete fourier transform (DFT) matrix $\mathbf{U}\in\mathbb{C}^{M\times M}$, which is composed of the array steering vectors of $M$ uniformly distributed orthogonal directions (beams) as $\mathbf{U}=[1,\mathbf{a}(\delta),\cdots,\mathbf{a}(\delta (M-1))]^{\rm{H}}$  with $\delta=\frac{1}{M}$ \cite{JTSP_HRobert_OverviewMmwave,TAP_JBrady_Beamspace}.
Therefore, the received signal $y_k$ at the $k$-th user can be expressed as
\begin{align}\label{eq_yk1}
y_k = \sqrt{\frac{\gamma}{K}}\mathbf{h}_k^{\rm{H}}\mathbf{U}^{\rm{H}}\mathbf{S}\mathbf{V}\mathbf{x}_d + n_k=\sqrt{\frac{\gamma}{K}}{(\mathbf{h}^b_k)}^{\rm{H}}\mathbf{S}\mathbf{V}\mathbf{x}_d + n_k,
\end{align}
where $\gamma$ is the transmit power, $\mathbf{x}_d=[x_{d,1},x_{d,2},\cdots,x_{d,K}]^{\rm{T}}\in\mathbb{C}^{K\times 1}$ is the data vector intended for $K$ users with normalized power $\rm{E}[|\mathbf{x}_d\mathbf{x}_d^{\rm{H}}|^2]=\mathbf{I}_K$, and the beamspace channel vector $\mathbf{h}_k^b\in\mathbb{C}^{M\times 1}$ can be expressed as
\begin{align}\label{eq_hkb}
\mathbf{h}_k^b=\mathbf{U}\mathbf{h}_k,
\end{align}
where the elements of $\mathbf{h}_k^b$ are channel coefficients associated with $M$ orthogonal beams. Note that $\mathbf{h}_k^b$ is sparse due to the limited scattering of mmWave channel \cite{TAP_JBrady_Beamspace}.
$\mathbf{S}\in\mathbb{C}^{M\times N_\text{RF}}$ is the beam selector with $N_\text{RF}$ denoting the number of RF chains at the BS. 
Each column of $\mathbf{S}$ has and only has one non-zero element `1', and the location of this non-zero element in one column of $\mathbf{S}$ should be different from that in another column. $n_k$ is the complex Gaussian noise at the $k$-th user with zero mean and unit variance.
We define the dimension-reduced equivalent channel $\mathbf{h}_k^e\in\mathbb{C}^{N_\text{RF}\times 1}$ as
\begin{align}\label{eq_hke}
\mathbf{h}_k^e=\mathbf{S}^{\rm{H}}\mathbf{h}^b_k,
\end{align}
thus we have
\begin{align}\label{eq_yk2}
y_k =\sqrt{\frac{\gamma}{K}}{(\mathbf{h}_k^e)}^{\rm{H}}\mathbf{V}\mathbf{x}_d + n_k,
\end{align}
where $\mathbf{V}=[\mathbf{v}_1,\mathbf{v}_2,\cdots,\mathbf{v}_K]\in\mathbb{C}^{N_\text{RF}\times K}$ is the zero-forcing (ZF) precoding matrix consisting of $K$ different unit-norm precoding vectors $\mathbf{v}_i\in\mathbb{C}^{N_\text{RF}\times 1}$,
which can be described as the normalized $i$-th column of ${(\mathbf{H}^e)}^\dagger$, i.e., $\mathbf{v}_i=\frac{{(\mathbf{H}^e)}^\dagger(:,i)}{\|{(\mathbf{H}^e)}^\dagger(:,i)\|}$, where $\mathbf{H}^e=[{\mathbf{h}_1^e},{\mathbf{h}_2^e},\cdots,{\mathbf{h}_K^e}] \in\mathbb{C}^{N_\text{RF}\times K}$.
Note that the equivalent channel (\ref{eq_hke}) is not Rayleigh distributed, and no dedicated codebook has been designed to quantize equivalent channel in the literature.
This gap can be filled by the proposed dimension-reduced subspace codebook in the next section.

\section{Proposed Dimension-Reduced Subspace Codebook}\label{S3}
In this section,
we briefly review the equivalent channel feedback in lens-based mmWave massive MIMO systems based on the classical codebook at first.
Then, we propose the dimension-reduced subspace codebook to quantize the equivalent channel vector better.
Finally, we provide the performance analysis of the proposed codebook by quantifying the required number of feedback bits.
\subsection{Equivalent Channel Feedback}\label{S3.1}
We assume that each user knows its equivalent channel vector $\mathbf{h}_k^e$ through channel estimation \cite{ICC_LYang_RFChannelEstimation}.
Then, the $k$-th user quantizes $\mathbf{h}_k^e$ to $B$ bits using a codebook $\mathcal{D}_k=\{\mathbf{d}_{k,1},\mathbf{d}_{k,2},\cdots,\mathbf{d}_{k,2^B}\}$. Considering the classical RVQ-based codebook, the quantization vector $\mathbf{d}_{k,i}\in\mathbb{C}^{N_\text{RF}\times 1}$ is randomly generated by selecting vectors independently from the uniform distribution on the complex $N_\text{RF}$-dimensional unit sphere \cite{TIT_NJindal_MIMOBroadcast}.
The number of required feedback bits $B$ should scale linearly with $N_\text{RF}$ to limit the system capacity loss within an acceptable level.

To realize channel feedback, the $k$-th user will find $\mathbf{d}_{k,F_k}$ that is the closest to $\mathbf{h}_k^e$,
where ``closeness'' is measured by the angle between two vectors.
That is, the index $F_k$ is computed according to
\begin{align}
F_k=\argmin_{i\in\{1,2,\cdots,2^B\}}\sin^2(\measuredangle(\mathbf{h}_k^e,\mathbf{d}_{k,i}))=\argmax_{i\in\{1,2,\cdots,2^B\}}|\mathbf{d}_{k,i}^{\rm{H}}\mathbf{\tilde{h}}_k^e|^2,
\end{align}
where $\mathbf{\tilde{h}}_k^e=\frac{\mathbf{h}_k^e}{\|\mathbf{h}_k^e\|}$ denotes the equivalent channel direction.
Since the equivalent channel magnitude $\|\mathbf{h}_k^e\|$ is just a scalar,
we follow the widely used assumption that the channel magnitude $\|\mathbf{h}_k^e\|$ can be fed back to the BS perfectly \cite{TIT_DJLove_Grassmannian,TIT_NJindal_MIMOBroadcast}.
The index $F_k$ can be fed back through $B$ bits,
then the BS can obtain the feedback equivalent channel vector $\hat{\mathbf{h}}_k^e=\|\mathbf{h}_k^e\|\mathbf{d}_{k,F_k}$.
However, since the equivalent channel vector $\mathbf{h}_k^e$ is not Rayleigh distributed, the classical RVQ-based codebook is not optimal in lens-based mmWave massive MIMO systems.
In the next subsection, we will propose the dimension-reduced subspace codebook to quantize this equivalent channel better.

\subsection{Proposed Dimension-Reduced Subspace Codebook}\label{S3.2}
Considering $\mathbf{h}_k=\mathbf{A}_k\mathbf{g}_k$ in (\ref{eq_hk2}),
we can find that the channel vector $\mathbf{h}_k$ is actually distributed on the column space of $\mathbf{A}_k\in\mathbb{C}^{M\times P_k}$,
which is formed by linear combination of $\mathbf{A}_k$'s column vectors.
Due to the severe propagation loss of mmWave signals \cite{JSSP_AAhmed_EstimationHybrid},
the number of paths is much smaller than the number of BS antenna, i.e. $P_k\ll M$.
Therefore, $P_k$ column vectors of $\mathbf{A}_k$ only form a subspace of the full $M$-dimensional space, which depends on $P_k$ different AoDs.
This subspace is called as \textit{channel subspace} in the rest of this paper.
Therefore, we firstly propose to generate high-dimensional vectors in the channel subspace.
Since the AoDs vary slower than path gains,
we assume that AoDs are slow varying \cite{TVT_HXie_Unified} and can be accurately estimated through beamspace channel estimation \cite{TWC_XGao_BeamspaceChannelEstimation}.
Then, we can generate the high-dimensional vector $\mathbf{c}_{k,i}\in\mathbb{C}^{M\times 1}$ as
\begin{align}\label{eq_cki}
\mathbf{c}_{k,i}=\mathbf{A}_k\mathbf{w}_{k,i},
\end{align}
where $\mathbf{w}_{k,i}\in\mathbb{C}^{P_k\times 1}$ is isotropically distributed on the complex $P_k$-dimensional unit sphere.
The high-dimensional vector $\mathbf{c}_{k,i}$ is distributed on the column space of $\mathbf{A}_k$, i.e., the channel subspace.

Then, based on the high-dimensional vector $\mathbf{c}_{k,i}$, we propose the dimension-reduced subspace codebook $\mathcal{D}_k=\{\mathbf{d}_{k,1},\mathbf{d}_{k,2},\cdots,\mathbf{d}_{k,2^B}\}$ to quantize the equivalent channel vector $\mathbf{h}_k^e$, where the quantization vector $\mathbf{d}_{k,i}\in\mathbb{C}^{N_\text{RF}\times 1}$ can be expressed as
\begin{align}\label{eq_dki1}
\mathbf{d}_{k,i}=\mathbf{S}^{\rm{H}}\mathbf{U}\mathbf{c}_{k,i},
\end{align}
where $\mathbf{U}$ and $\mathbf{S}$ are used to reflect the function of lens and beam selector in lens-based mmWave massive MIMO systems, respectively.
Note that to ensure the unit-norm vector requirement of $\mathbf{d}_{k,i}$, $\mathbf{d}_{k,i}$ should be normalized as
\begin{align} \label{eq_dki2}
\vspace{-1mm}
\mathbf{d}_{k,i}=\frac{\mathbf{d}_{k,i}}{\|\mathbf{d}_{k,i}\|}.
\vspace{-1mm}
\end{align}
Considering (\ref{eq_hkb}) and (\ref{eq_hke}), we can find that the quantization vector $\mathbf{d}_{k,i}$ in the proposed dimension-reduced subspace codebook $\mathcal{D}_k$ has the same distribution as the equivalent channel vector $\mathbf{h}_k^e$.
Thus, the proposed codebook is expected to enjoy better quantization performance than the classical RVQ-based codebook without considering the special channel characteristics of lens-based mmWave massive MIMO systems.
Such intuitive conclusion will be verified by the theoretical analysis of the proposed codebook in terms of the required number of feedback bits in the next subsection.

\subsection{Performance Analysis of The Proposed Codebook}\label{S3.3}
After the channel feedback based on the proposed dimension-reduced subspace codebook, the BS can perform downlink precoding based on the feedback equivalent channel matrix $\mathbf{\hat{H}}^e=[\mathbf{\hat{h}}_1^e,\mathbf{\hat{h}}_2^e,\cdots,\mathbf{\hat{h}}_K^e]\in\mathbb{C}^{N_{\text{RF}}\times K}$.
We can rewrite (\ref{eq_yk2}) as
 \begin{align}
y_k=\sqrt{\frac{\gamma}{K}}{(\mathbf{h}_k^e)}^{\rm{H}}\mathbf{v}_kx_{d,k}+\sqrt{\frac{\gamma}{K}}\sum_{i=1,i\neq k}^{K}{(\mathbf{h}_k^e)}^{\rm{H}}\mathbf{v}_ix_{d,i}+n_k,
\end{align}
where the second term denotes the multi-user interference. Thus, the per user rate $R$ with limited channel feedback based on the proposed dimension-reduced subspace codebook is
 \begin{align}
R=\rm{E}\left[ \log_2\left( 1+\frac{\frac{\gamma}{K}|{(\mathbf{h}_k^e)}^{\rm{H}}\mathbf{v}_k|^2}{1+\frac{\gamma}{K}\sum_{i=1,i\neq k}^{K}|{(\mathbf{h}_k^e)}^{\rm{H}}\mathbf{v}_i|^2}\right) \right] .
\end{align}

For the ideal case of perfect CSI at the BS, i.e., $\mathbf{\hat{H}}^e=\mathbf{H}^e$,
the ZF precoding vector $\mathbf{v}_{\text{ideal},i}$ is obtained as the normalized $i$-th column of ${(\mathbf{H}^e)}^\dagger$.
Thus, $\mathbf{v}_{\text{ideal},i}$ is orthogonal to the $k$-th user's equivalent channel vector $\mathbf{h}_k^e$ for any $i\neq k$,
which means that we can obtain the ideal per user rate as
 \begin{align}
R_\text{ideal}=\rm{E}\left[ \log_2\left( 1+\frac{\gamma}{K}|{(\mathbf{h}_k^e)}^{\rm{H}}\mathbf{v}_{\text{ideal},k}|^2\right) \right] .
\end{align}
We define the rate gap $\Delta R(\gamma)$ as the difference between the per user rate with ideal CSI and that with limited channel feedback:
\begin{align}
 \vspace{-1mm}
\Delta R(\gamma) &= R_\text{ideal}-R.
 \vspace{-1mm}
\end{align}
Following the results from \cite{TIT_NJindal_MIMOBroadcast}, the rate gap $\Delta R(\gamma)$ can be upper bounded as
\begin{align}\label{eq_DR}
\Delta R(\gamma)\!\leq\! \log_2\left(\!1\!+\!\frac{\gamma(K\!-\!1)}{K}\!\rm{E}\!\left[ \|\mathbf{h}_k^e\|^2\right]\!\rm{E}\!\left[ \sin^2(\measuredangle(\tilde{\mathbf{h}}_k^e,\hat{\mathbf{h}}_k^e))\right] \right) .
\end{align}

We can observe from (\ref{eq_DR}) that the rate gap $\Delta R(\gamma)$ depends on the equivalent channel norm $\|\mathbf{h}_k^e\|$ and the quantization error $\sin^2(\measuredangle(\tilde{\mathbf{h}}_k^e,\hat{\mathbf{h}}_k^e))$. In the rest of this paper, we can omit the subscript $k$ without loss of generality, where $\mathbf{h}^e$, $\mathbf{A}$, and $P$ denote the equivalent channel, steering matrix, and number of resolvable paths, respectively. 
Next, we will discuss the quantization error $\rm{E}\left[ \sin^2(\measuredangle(\tilde{\mathbf{h}}^e,\hat{\mathbf{h}}^e))\right]$ in (\ref{eq_DR}).
Before that, we will prove Lemma 1 and Lemma 2 below which will be useful for the analysis of the quantization error.

\textit{\textbf{Lemma 1}}: Denoting $\mathbf{T}=\mathbf{S}^{\rm{H}}\mathbf{U}\mathbf{A}$, we have $\mathbf{T}^{\rm{H}}\mathbf{T}=\mathbf{I}_P$.
\begin{IEEEproof}
Since the $i$-th row vector $\mathbf{a}(\delta (j-1))^{\rm{H}}$ of $\mathbf{U}$ and the $j$-th column vector $\mathbf{a}(\psi_{i})$ of $\mathbf{A}$ is asymptotically orthogonal if $\psi_{i}\neq \delta (j-1)$ when $M\rightarrow\infty$,
$\mathbf{U}\mathbf{A}$ has and only has one non-zero element ``1" in each column, and the location of this non-zero element in one column is different from that in another column.
Generally, the beam selector $\mathbf{S}$ is designed to choose $N_\text{RF}$ rows of $\mathbf{U}\mathbf{A}$ which includes non-zero elements \cite{CL_XGao_BeamSelection}. Note that for the multi-user scenario, $N_\text{RF}$ is usually larger than $P$. 
Thus, $\mathbf{T}\in\mathbb{C}^{N_\text{RF}\times P}$ has and only has one non-zero element ``1" in each column, and the locations of non-zero elements are different for different columns.
Therefore, we have $\mathbf{T}^{\rm{H}}\mathbf{T}=\mathbf{I}_P$.
\end{IEEEproof}

\textit{\textbf{Lemma 2}}: $\|\mathbf{h}^e\|=\|\mathbf{S}^{\rm{H}}\mathbf{U}\mathbf{A}\mathbf{g}\|=\|\mathbf{g}\|$.
\begin{IEEEproof}
Combining (\ref{eq_hk2}), (\ref{eq_hkb}) and (\ref{eq_hke}), we have
\begin{align}\label{eq_e}
\mathbf{h}^e=\mathbf{S}^{\rm{H}}\mathbf{U}\mathbf{A}\mathbf{g}.
\end{align}
By denoting $\mathbf{S}^{\rm{H}}\mathbf{U}\mathbf{A}=\mathbf{T}=[\mathbf{t}_1,\mathbf{t}_2,\cdots,\mathbf{t}_P]$, we have
\begin{align} \label{eq_ci_norm}
\|\mathbf{h}^e\|&=\|\mathbf{T}\mathbf{g}\|=\|\Sigma_{p=1}^P\mathbf{t}_pg_{p}\| \\\nonumber
&\overset{(a)}{=}\Sigma_{p=1}^P\|\mathbf{t}_pg_{p}\| \overset{(b)}{=}\Sigma_{p=1}^P|g_{p}|=\|\mathbf{g}\|,
\end{align}
where (a) is true due to the orthogonality among column vectors $\mathbf{t}_p$ of $\mathbf{T}$ (see Lemma 1), and (b) is true due to $\|\mathbf{t}_p\|=1$ (see Lemma 1).
\end{IEEEproof}
Now, we will provide an upper bound of the quantization error in (\ref{eq_DR}) in the following Lemma 3.

\textit{\textbf{Lemma 3}}: The quantization error $\rm{E}\left[ \sin^2(\measuredangle(\tilde{\mathbf{h}}^e,\hat{\mathbf{h}}^e))\right]$ of equivalent channel vector can be upper bounded as
\begin{align}\label{eq_Esin}
\rm{E}\left[ \sin^2(\measuredangle(\tilde{\mathbf{h}}^e,\hat{\mathbf{h}}^e))\right] <2^{-\frac{B}{P-1}}.
\end{align}
\begin{IEEEproof}
Since $\|\tilde{\mathbf{h}}^e\|=1$, $\hat{\mathbf{h}}^e=\|\mathbf{h}^e\|\mathbf{d}_F$, and $\|\mathbf{d}_F\|=1$, we have
\begin{align}\label{sin2e}
\rm{E}\left[ \sin^2(\measuredangle(\tilde{\mathbf{h}}^e,\hat{\mathbf{h}}^e))\right] & = 1-\rm{E}\left[ \cos^2(\measuredangle(\tilde{\mathbf{h}}^e,\hat{\mathbf{h}}^e))\right]  \\\nonumber
&=1-\rm{E}\left[ |\mathbf{d}_F^{\rm{H}}{\tilde{\mathbf{h}}^e}|^2\right].
\end{align}
By denoting $\tilde{\mathbf{g}}=\frac{\mathbf{g}}{\|\mathbf{g}\|}$ and using Lemma 2,
we can get $\tilde{\mathbf{h}}^e=\frac{\mathbf{T}\mathbf{g}}{\|\mathbf{h}^e\|}=\mathbf{T}\tilde{\mathbf{g}}$.
Considering (\ref{eq_dki1}) and using $\|\mathbf{w}_F\|=1$, we have $\mathbf{d}_F=\mathbf{T}\mathbf{w}_F$.
Therefore,
\begin{align}\label{ed}
\rm{E}\left[ |\mathbf{d}_F^{\rm{H}}{\tilde{\mathbf{h}}^e}|^2\right] =\rm{E}\left[ |\mathbf{w}_F^{\rm{H}}\mathbf{T}^{\rm{H}}\mathbf{T}\tilde{\mathbf{g}}|^2\right] \overset{(a)}{=}\rm{E}\left[ |\mathbf{w}_F^{\rm{H}}\tilde{\mathbf{g}}|^2\right],
\end{align}
where (a) follows from the results $\mathbf{T}^{\rm{H}}\mathbf{T}= \mathbf{I}_P$ in Lemma 1.
Since $\mathbf{w}_F$ and $\tilde{\mathbf{g}}$ are isotropically distributed $P$-dimensional vectors,
$\rm{E}\left[ |\mathbf{w}_F^{\rm{H}}\tilde{\mathbf{g}}|^2\right]$ can be lower bounded as \cite{TIT_NJindal_MIMOBroadcast}:
\begin{align}\label{glowbound}
\rm{E}\left[ |\mathbf{w}_F^{\rm{H}}\tilde{\mathbf{g}}|^2\right] >1-2^{-\frac{B}{P-1}}.
\end{align}
Combine (\ref{sin2e}), (\ref{ed}), and (\ref{glowbound}), we can obtain (\ref{eq_Esin}).
\end{IEEEproof}

Substituting (\ref{eq_Esin}) into (\ref{eq_DR}), we have
\begin{align}\label{delatR4}
\Delta R(\gamma)&\leq \log_2\left( 1+\frac{\gamma(K-1)}{K}\rm{E}\left[ \|\mathbf{h}^e\|^2\right] 2^{-\frac{B}{P-1}}\right) .
\end{align}
Let the rate gap $\Delta R(\gamma)\leq 1$ bps/Hz \cite{TIT_NJindal_MIMOBroadcast}, then the number of feedback bits $B$ should scale according to
\begin{align}\label{eq_delatR_UpBound}
B \geq \frac{P-1}{3}\text{SNR}+(P-1)\log_2(K-1),
\end{align}
where $\text{SNR}=10\log_{10}\frac{\gamma}{K}\rm{E}\left[ \|\mathbf{h}^e\|^2\right] $ is the signal-to-noise-ratio (SNR) at the receiver.
We can observe from (\ref{eq_delatR_UpBound}) that the slope of the required number of feedback bits $B$ is $P-1$ when $\text{SNR}$ increases.
In other words, the required number of feedback bits $B$ only scales linearly with $P-1$ to maintain a constant rate gap.
Since $P\ll N_\text{RF}$,
the proposed dimension-reduced subspace codebook require much less feedback overhead than the classical RVQ-based codebook where the number of feedback bits scales linearly with $N_\text{RF}$. In next section, we will verify our analysis through simulations.

\section{Simulation Results}\label{S4}
A simulation study was carried out to verify the performance of the proposed dimension-reduced subspace codebook in lens-based mmWave massive MIMO systems.
The key system parameters are set as:
 1) the number of BS antennas, the number of RF chains, the number of users and the number of resolvable paths are $(M,N_\text{RF}, K,P)=(128,24,8,3)$;
 2) the AoDs are randomly chosen from uniform distribution $\mathbb{U}[-\frac{1}{2}\pi, \frac{1}{2}\pi]$;
 3) the number of feedback bits $B=\frac{P-1}{3}\text{SNR}+ (P-1)\log_2(K-1)$ for all considered codebooks for fair comparison.

\begin{figure}
\vspace{-2mm}
\center{\includegraphics[width=0.5\textwidth]{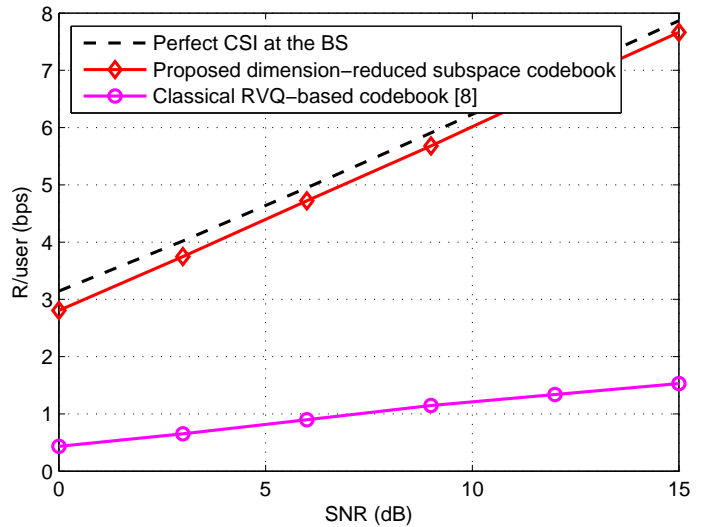}}
\vspace{-3mm}
\caption{Per user rate of the ideal case of perfect CSI and the practical case of limited channel feedback.}
\label{Fig1}
\end{figure}

Fig.~\ref{Fig1} shows per user rate of the ideal case of perfect CSI at the BS and the practical case of limited channel feedback,
where the proposed codebook and the classical RVQ-based codebook are compared.
We can observe that the rate gap between the ideal case of perfect CSI at the BS and the practical case of limited channel feedback using the proposed codebook remains constant when SNR increases,
which is consistent with our theoretical analysis in Section \ref{S3.3}.
On the contrary, for the classical RVQ-based codebook,
the rate gap increases with the SNR.
We can also find that the proposed codebook outperforms the classical RVQ-based codebook.

\section{Conclusions}\label{S5}
In this paper, we have investigated the problem of codebook design for equivalent channel feedback in lens-based mmWave massive MIMO systems for the first time.
Specifically, we firstly proposed to utilize the limited scattering property of mmWave channels to generate the high-dimensional vectors in the channel subspace.
After that, we further proposed the dimension-reduced subspace codebook to quantize the equivalent channel vector according to the function of lens and beam selector.
We also provided the performance analysis of the proposed dimension-reduced subspace codebook.
Simulation results verified that the proposed codebook has better performance than the traditional codebook.

\bibliographystyle{IEEEtran}
\bibliography{IEEEabrv,Gao1Ref}

\end{document}